\pdfoutput=1
\documentclass[11pt]{article}
\usepackage{times}
\usepackage{latexsym}

\usepackage{graphicx}
\usepackage{amsmath}
\usepackage{amsfonts}
\usepackage{multirow}
\usepackage{float}
\usepackage{booktabs}
\usepackage{makecell}
\usepackage{multirow}

\usepackage[]{acl}

\usepackage{times}
\usepackage{latexsym}

\usepackage[T1]{fontenc}
\usepackage[utf8]{inputenc}

\usepackage{microtype}

\title{SpeechT5: Unified-Modal Encoder-Decoder Pre-Training for \\ Spoken Language Processing}

\author{
Junyi Ao$^{1,2,}$\thanks{\ \ Equal contribution. Work  is done by the first two authors during internship at Microsoft Research Asia. Correspondence to:  Long Zhou (lozhou@microsoft.com)} , Rui Wang$^{3,\ast}$, Long Zhou$^{4,\ast}$, Chengyi Wang$^4$, Shuo Ren$^4$, \\
{\bf Yu Wu}$^4${\bf,} {\bf Shujie Liu}$^4${\bf,} {\bf Tom Ko}$^1${\bf,} {\bf Qing Li}$^2${\bf,} {\bf Yu Zhang}$^{1,5}${\bf,} {\bf Zhihua Wei}$^3${\bf,}\\
{\bf Yao Qian}$^4${\bf,} {\bf Jinyu Li}$^4${\bf,} {\bf Furu Wei}$^4$\\
$^1$Department of Computer Science and Engineering, \\
Southern University of Science and Technology\\
$^2$Department of Computing, The Hong Kong Polytechnic University\\
$^3$Department of Computer Science and Technology, Tongji University\\
$^4$Microsoft 
$^5$Peng Cheng Laboratory\\
}

\date{}

\begin{document}
\maketitle
\begin{abstract}

Motivated by the success of T5 (Text-To-Text Transfer Transformer) in pre-trained natural language processing models, we propose a unified-modal SpeechT5 framework that explores the encoder-decoder pre-training for self-supervised speech/text representation learning.
The SpeechT5 framework consists of a shared encoder-decoder network and six modal-specific (speech/text) pre/post-nets. 
After preprocessing the input speech/text through the pre-nets, the shared encoder-decoder network models the sequence-to-sequence transformation,
and then the post-nets generate the output in the speech/text modality based on the output of the decoder.
Leveraging large-scale unlabeled speech and text data, we pre-train SpeechT5 to learn a unified-modal representation, hoping to improve the modeling capability for both speech and text.
To align the textual and speech information into this unified semantic space, we propose a cross-modal vector quantization approach that randomly mixes up speech/text states with latent units as the interface between encoder and decoder.
Extensive evaluations show the superiority of the proposed SpeechT5 framework on a wide variety of spoken language processing tasks, including automatic speech recognition, speech synthesis, speech translation, voice conversion, speech enhancement, and speaker identification.
We release our code and model at
\url{https://github.com/microsoft/SpeechT5}.

\end{abstract}

\section{Introduction}

%Starting with ELMo \cite{peters2018deep} and BERT \cite{devlin2018bert}, substantial work has shown that pre-trained models can bring significant improvements in various tasks, including natural language processing (NLP), image recognition, and speech processing \cite{radford2019language,lample2019cross,yang2019xlnet,dong2019unified,lewis2019bart,bao2021beit,baevski2020wav2vec}.
%It is becoming a new principle to solve problems by pre-training a universal model with self-supervision tasks on a large amount of unlabeled data to learn universal representations, followed with a fine-tuning stage on the down-streaming tasks.

% Particularly, ``Text-To-Text Transfer Transformer'' (T5) \cite{raffel2019exploring} leverages a unified text-to-text framework and achieves state-of-the-art results on a wide variety of NLP tasks, including machine translation, question answering, sentiment classification, and so on. The basic idea of T5 is to formulate every NLP problem as a ``text-to-text'' problem, and employ transfer learning to boost the performance of downstream tasks.

\begin{figure}[t]
	\centering
	\setlength{\abovecaptionskip}{3pt}
    \setlength{\belowcaptionskip}{-0pt}
	\includegraphics[width=7.4cm]{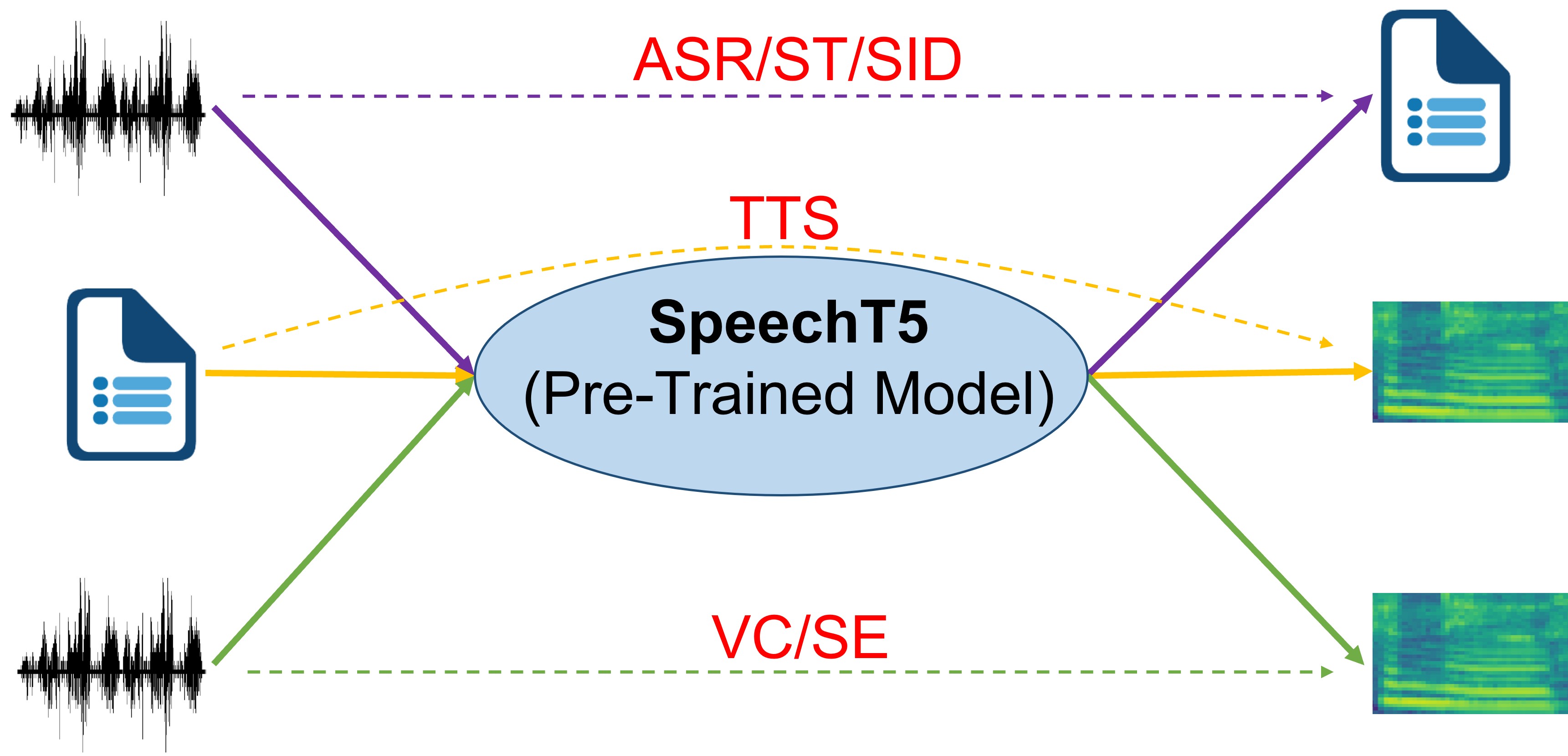}
	\caption{An illustration of the SpeechT5 framework, which treats spoken language processing tasks as a speech/text to speech/text format, including automatic speech recognition (ASR), speech translation (ST), speech identification (SID), text to speech (TTS),  voice conversion (VC), and speech enhancement (SE).
	}\label{speecht5}
\vspace{-12pt}\end{figure}%

Starting with ELMo \cite{peters2018deep} and BERT \cite{devlin2018bert}, substantial work has shown that pre-trained models can significantly improve in various natural language processing (NLP) tasks \cite{radford2019language,lample2019cross,yang2019xlnet,dong2019unified,lewis2019bart}.
Following the pre-training techniques in NLP, self-supervised speech representation learning has also been investigated and shown promising results, benefiting from richly learned representations \cite{chung2018speech2vec,chuang2019speechbert,song2019speech,baevski2020wav2vec,wang2021unispeech,hsu2021hubert,chung2021w2v}, such as wav2vec 2.0 \cite{baevski2020wav2vec} and HuBERT \cite{hsu2021hubert}.

However, previous speech pre-training work suffers from two problems: (1) most of them learn the speech representation with only unlabeled speech data but ignore the importance of textual data to spoken language tasks (e.g., automatic speech recognition) which require the modality transformation; (2) most of these models solely rely on a pre-trained speech encoder for various downstream tasks, leaving the decoder not pre-trained for the sequence-to-sequence generation tasks. How to design a unified encoder-decoder model that can take advantage of both unlabeled speech and text data to improve various spoken language processing tasks is not well explored.

Inspired by the T5 method \cite{raffel2019exploring}, we attempt to formulate each spoken language processing task as a speech/text to speech/text problem via an encoder-decoder framework, which enables us to use the same pre-trained model with bimodal data across diverse tasks, as shown in Figure \ref{speecht5}.
To achieve this, we propose a unified-modal pre-training framework, SpeechT5, containing an encoder-decoder backbone network and modal-specific pre/post-nets. 
With the pre-nets, the input speech/text is embedded in a shared space, and the encoder-decoder backbone network models the sequence-to-sequence conversion, from which the model-specific post-nets generate the speech/text output.
Particularly, SpeechT5 is mainly pre-trained with a denoising sequence-to-sequence method by leveraging large-scale unlabeled text and speech corpus.
To align the textual and acoustic information into a unified semantic space, the proposed SpeechT5 model (1) maps text and speech representations into a shared vector quantization space, and (2) randomly mixes up the quantized latent representations and the contextual states, which can better guide the quantizer to learn the cross-modal features.

We fine-tune SpeechT5 on a wide variety of downstream spoken language processing tasks, including automatic speech recognition (ASR), text-to-speech (TTS), speech translation (ST), voice conversion (VC), speech enhancement (SE), and speaker identification (SID).
Massive experiments show that the proposed SpeechT5 model achieves a significant improvement on these spoken language processing tasks compared with the state-of-the-art baselines.
Specifically, the proposed SpeechT5 outperforms wav2vec 2.0 \cite{baevski2020wav2vec} and HuBERT \cite{hsu2021hubert} with the \textsc{Base} model on the ASR task and also performs better than the state-of-the-art voice Transformer network \cite{huang2021pretraining} on the VC task.
Besides, SpeechT5 is significantly superior to SpeechNet \cite{chen2021speechnet} and pre-trained models from SUPERB \cite{yang2021superb} and achieves the state-of-the-art performance (i.e., 96.49\%) on the SID task.
We further provide an empirical comparison of the pre-training tasks and modules, and the ablation study demonstrates the effectiveness of the proposed joint speech-text pre-training method.

% Specifically, the proposed SpeechT5 method performs better than the state-of-the-art voice Transformer network (VTN) \cite{huang2021pretraining} on the VC task, and achieves the state-of-the-art result of 90.97\%. It also outperforms SpeechNet \cite{chen2021speechnet} and pre-trained models such as SUPERB \cite{yang2021superb} on the SID task.
% Besides, SpeechT5 also obtains a gain of about 10.0\% and 6.5\% than the encoder-decoder based ASR model (i.e., Fairseq \cite{ott2019fairseq}) and some baseline models, respectively, on the ASR task, and obtains significant improvements over the strong Transformer TTS model \cite{li2019neural} by 13.4\% and 5.8\% in terms of the word error rate and mean option score (MOS) on the TTS task.

The contributions of this paper are summarized as follows.

\begin{itemize}
\item To the best of our knowledge, this is the first work to investigate a unified encoder-decoder framework for various spoken language processing tasks.
\item We propose a cross-modal vector quantization approach, which learns the implicit alignment between acoustic and textual representation with large-scale unlabeled speech and text data.
\item Extensive experiments on spoken language processing tasks demonstrate the effectiveness and superiority of the proposed SpeechT5 model.
\end{itemize}

\begin{figure*}[!ht]
	\centering
	\includegraphics[width=16cm]{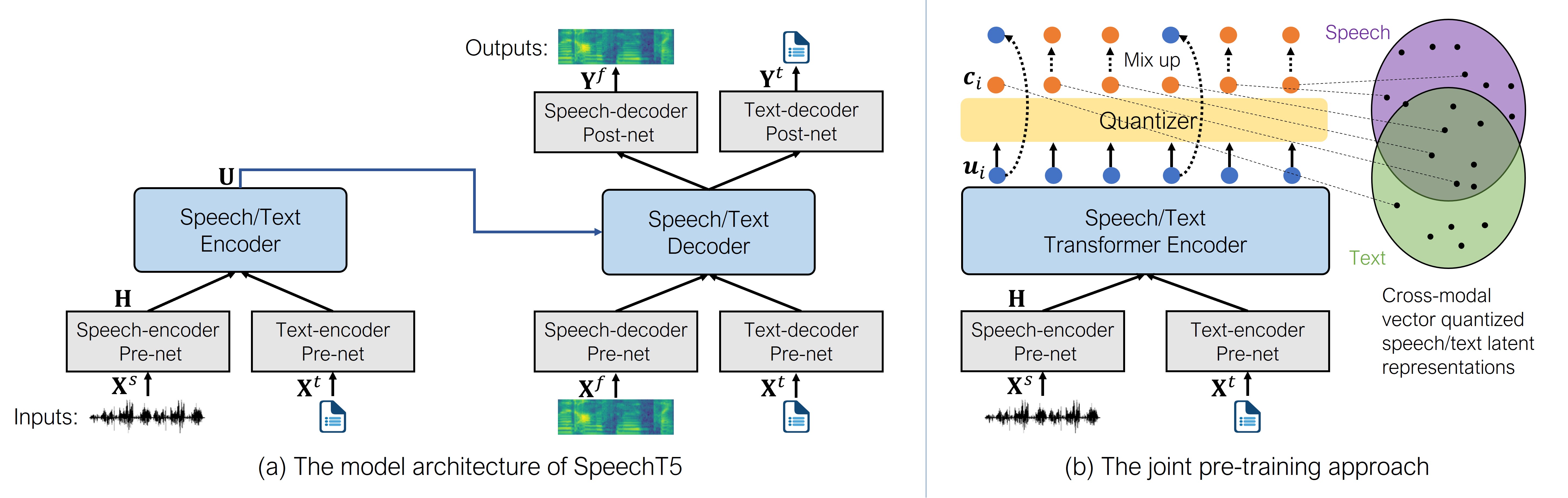}
	\caption{(a) The model architecture of SpeechT5, which contains an encoder-decoder module and six modal-specific pre/post-nets. Most spoken language processing tasks can be learned by concatenating the encoder-decoder module and the corresponding pre-net and post-net. (b) By sharing discrete tokens across modalities, the joint pre-training approach builds bridges between speech and text. Hidden states and latent units are mixed up and used as the inputs of the cross-attention module in the decoder.
	%The shared quantization module over encoder representations produces cross-modality quantized latent speech/text units.  The model learns to share discrete tokens across modalities, creating bridges between speech and text.
	}\label{architecture}
\end{figure*}

\section{SpeechT5}

In this section, we propose SpeechT5, a unified-modal framework for learning joint contextual representations for speech and text data via a shared encoder-decoder structure.

\subsection{Model Architecture}
\label{sec_architecture}

% lack of motivation behind the model design
%All spoken language processing tasks take speech or text as the input or output. 
Figure \ref{architecture}(a) shows the model architecture of the proposed SpeechT5 model. It consists of an encoder-decoder module and six modal-specific pre/post-nets.
The pre-nets convert the input speech $\mathbf{X}^s \in \mathcal{D}^s$ or text $\mathbf{X}^t \in \mathcal{D}^t$ to a unified space of hidden representations and then feed them into the shared encoder-decoder to perform the sequence-to-sequence conversion. Finally, the post-nets generate the output in the speech or text modality, based on the decoder output.

\paragraph{Input/Output Representations}
To train a single model for a diverse set of spoken language processing tasks, we formulate them as ``speech/text to speech/text'' tasks, where the model is fed with speech/text as the input and generates the corresponding output in the speech/text format.
Specifically, a text is split into a sequence of characters  $\mathbf{X}^t = (\mathbf{x}^t_1,..., \mathbf{x}^t_{N^t})$ as the input and output. For speech modality, the raw waveform $\mathbf{X}^s  = (\mathbf{x}^s_1,..., \mathbf{x}^s_{N^s})$ is used as the input, and a sequence of the log Mel-filterbank features $\mathbf{X}^f  = (\mathbf{x}^f_1,..., \mathbf{x}^f_{N^f})$ extracted from raw audio using librosa tool{\footnote[1]{https://librosa.org/doc/latest/index.html.}} is adopted as the target output.
A vocoder \cite{kong2020hifi} is leveraged to generate the final waveform from the generated features. 

%For text, we split the text into a sequence of tokens using a unigram language model \cite{kudo2018subword}.
% To specify which task the model should perform, we add a task-specific prefix to the original input sequence before feeding it to the model.

\paragraph{Encoder-Decoder Backbone} The Transformer encoder-decoder model \cite{vaswani2017attention} is used as the backbone network of SpeechT5. 
Please refer to \citet{vaswani2017attention} for more details.
% The encoder consists of a stack of blocks, each of which comprises two components: a self-attention layer and a feed-forward network (FFN) following the self-attention layer.
% The decoder has a similar architecture to the encoder except that it includes a cross-attention mechanism after each self-attention layer that attends to the output of the encoder, and an autoregressive or causal self-attention is used to only attend to the past outputs.
%Layer normalization \cite{ba2016layer} and residual connection \cite{he2016deep} are applied to the input of each subcomponent.
%Dropout \cite{srivastava2014dropout} is applied within the feed-forward network, on the skip connection, on the attention weights, and at the input and output of the entire stack.
We employ the relative position embedding \cite{shaw-etal-2018-self} to help capture the relative position differences between elements in the input. Specifically, we only add the relative position embedding to the dot-product weights of the self-attention.

\paragraph{Speech Pre/Post-Net}
%There are some differences between the speech-encoder pre-net and speech-decoder pre-net.
The convolutional feature extractor of wav2vec 2.0 \cite{baevski2020wav2vec} serves as the speech-encoder pre-net to downsample raw waveform $\mathbf{X}^s$ and produce a sequence of a speech utterance $\mathbf{H}=(\mathbf{h}_1,...,\mathbf{h}_{N^{h}})$.
The speech-decoder pre-net is a neural network composed of three fully connected layers with the ReLU activation, fed with the log Mel-filterbank $\mathbf{X}^f$. 
To support multi-speaker TTS and VC, the speaker embedding extracted with the x-vector \cite{Snyder2018} is concatenated with the output of the speech-decoder pre-net followed by a linear layer.
% For the speech-decoder post-net, we use two different linear projections to predict the log Mel-filterbanks and the stop token, respectively, and use five 1-dimensional convolutional layers
%with the batch normalization and Tanh activation 
% to produce a residual to refine the log Mel-filterbank.
The speech-decoder post-net consists of two modules.
The first module uses a linear layer fed with the decoder output to predict the log Mel-filterbank $\mathbf{Y}^f = (\mathbf{y}_1^f,..., \mathbf{y}_{N^f}^f)$, followed by five 1-dimensional convolutional layers to produce a residual to refine the predicted $\mathbf{Y}^{f}$.
Another linear module is added to project the decoder output to a scalar for predicting the stop token.

\paragraph{Text Pre/Post-Net}
We use shared embeddings as the text-encoder pre-net and text-decoder pre/post-nets. The pre-net transforms a token index into an embedding vector. The post-net transforms the hidden state into the probability distribution of tokens, normalized by the softmax function. 
%There is a shift in the input text of the decoder for the auto-regressive generation. 
%During the inference, the decoder uses its own past predictions to predict the next token.

\subsection{Pre-Training}
\label{sec_pretraining}

% The proposed SpeechT5 model can be pre-trained with large-scale collections of unlabeled speech and text corpus, and the textual and acoustic information can be aligned into a unified semantic space via the proposed joint pre-training method.
The proposed SpeechT5 model can be pre-trained with large-scale collections of unlabeled speech and text corpus. The proposed joint pre-training method can align the textual and acoustic information into a unified semantic space.

\paragraph{Speech Pre-Training}

Leveraging unlabeled speech data $\mathcal{D}^s$ to learn general speech representations for both classification and generation tasks, SpeechT5 is trained with two types of tasks: bidirectional masked prediction and sequence-to-sequence generation. 
%To that end, our model is trained as a unified encoder-decoder model with sequence-to-sequence generation task. 

%Formally, the input to the speech module is a sequence of 80-dimensional log Mel-filterbank $X=(x_1,...,x_n)$.
%The speech module, which consists of a speech-encoder pre-net and a Transformer encoder, produces hidden representations $S=(s_1,...,s_n)$.
% We can apply any type of corruption into the input of Transformer encoder.
% Concretely, we use span mask strategies, where $p$\% of timesteps are randomly selected as start indices, and spans of $l$ steps are masked.
% Formally, $\mathbf{H}=(\mathbf{h}_1,...,\mathbf{h}_N)$ denotes a speech utterance of $N$ frames, which is transformed from raw audio $\mathbf{x}^s \in \mathcal{D}^{S}$ via the speech-encoder pre-net.
% Taking masked $\mathbf{H}$ as the input, the Transformer encoder produces hidden representations $\mathbf{U}=(\mathbf{u}_1,...,\mathbf{u}_N)$.
Following HuBERT \cite{hsu2021hubert}, the bidirectional masked prediction leverages a masked language model similar to BERT \cite{devlin2018bert} for the encoder, in which an acoustic unit discovery model provides the frame-level targets $\mathbf{Z}=(\mathbf{z}_1,...,\mathbf{z}_{N^{h}})${\footnote[2]{The target labels are generated by clustering outputs of the 6-th Transformer layer in the first iteration HuBERT \textsc{Base} model via the $k$-means clustering method with 500 clusters.}}. 
Specifically, we apply span mask strategies to the output $\mathbf{H}$ from speech-encoder pre-net, where 8\% of timesteps are randomly selected as start indices, and spans of 10 steps are masked. 
The Transformer encoder takes masked $\mathbf{H}$ as the input and produces hidden representations $\mathbf{U}=(\mathbf{u}_1,...,\mathbf{u}_{N^{h}})$.
Based on these hidden representations, the cross-entropy loss is computed over masked timesteps as
\begin{equation}
%\begin{aligned}
    \mathcal{L}_{mlm}^{s} = \sum_{n\in \mathcal{M}} \log p (\mathbf{z}_n|\hat{\mathbf{H}}, n),
%\end{aligned}
\end{equation}
where $\hat{\mathbf{H}}$ denotes the masked version of $\mathbf{H}$, $\mathcal{M}$ denotes the set of masked timesteps, and $\mathbf{z}_n$ denotes the frame-level target at timestep $n$ from $\mathbf{Z}$.

Furthermore, we propose to reconstruct the original speech via a sequence-to-sequence generation task, given the randomly masked input as introduced in bidirectional masked prediction.
%The decoder is auto-regressive in that the output of the encoder $\mathbf{U}$ and the previously generated features $\mathbf{y}_{1:t-1}$ are considered when decoding the current log Mel-filterbank feature $\mathbf{y}_t$.
Following seq2seq TTS models \cite{li2019neural}, we enforce the corresponding predicted output $ \mathbf{Y}^f$, which is generated through the speech-decoder pre-net, Transformer decoder, and speech-decoder post-net, to be close to the original $\mathbf{X}^f$ by minimizing their $L_1$-distance as
\begin{equation}
\begin{aligned}
    \mathcal{L}_1^{s} &= \sum_{n=1}^{N^f} \| \mathbf{y}_n^{f} - \mathbf{x}_n^{f}  \|_1, %\\
    % \mathcal{L}_2^{s} &= \sum_{i=1}^n ||y_i - x_i||_2.
\end{aligned}
\end{equation}
where $\mathbf{x}_n^{f}$ denotes $n$-th an 80-dimensional log Mel-filterbank from $\mathbf{X}^{f}$.
%extracted from raw waveform $\mathbf{x}_n^s \in \mathbf{X}^s$.
Besides, we use the binary cross-entropy (BCE) loss $\mathcal{L}_{bce}^{s}$ for the stop token.
%To address the imbalance problem between stop tokens and normal tokens, we impose a positive weight on the tail positive stop token when calculating the BCE loss.

\paragraph{Text Pre-Training}

%The language module aims to offer contextual understanding and generation.
With unlabeled text data $\mathcal{D}^t$, SpeechT5 is trained to reconstruct the model output $\mathbf{Y}^t = (\mathbf{y}_1^t,..., \mathbf{y}_{N^t}^t)$ to the original text $\mathbf{X}^t$, using the corrupted text $\hat{\mathbf{X}}^t =(\hat{x}_1^t,...,\hat{x}_{M}^t)$ as the input generated with a mask-based noising function.
% We follow BART \cite{lewis2019bart} to mask the input and generate the target sequence instead of the T5 strategy \cite{raffel2019exploring}, as we found that the former is better on most experiments (as shown in Appendix \ref{sec:appendix_mask}).
%We follow BART \cite{lewis2019bart} to mask the input by the text infilling approach and predict the original sequence.{\footnote[2]{We conducted experiments to campare the BART \cite{lewis2019bart} and T5 \cite{raffel2019exploring} mask strategies, which can be found in Appendix \ref{sec:appendix_mask}. }}
Following the text infilling approach in BART{\footnote[3]{We conducted experiments to compare the BART \cite{lewis2019bart} and T5 \cite{raffel2019exploring} mask strategies, which can be found in Appendix \ref{sec:appendix_mask}. }} \cite{lewis2019bart}, we randomly sample 30\% of text spans to mask, where the span length of text spans draws from a Poisson distribution ($\lambda = 3.5$), and each span is replaced with a single mask token.
%One mask token is used to replace each span.
%In the BART-style, the model aims to reconstruct the original text $\mathbf{X} \in \mathcal{D}^\text{T}$ from the noisy source text $\hat{\mathbf{X}}$.
%However, in the T5-style, all selected fragments are removed from the text and concatenated as the target sequence, while the remaining parts are concatenated as the source sequence.
Formally, SpeechT5, including text-encoder pre-net, encoder-decoder model, and text-decoder pre/post nets, is optimized to generate the original sequence with the maximum likelihood estimation as
%is trained to generate the original sequence $\mathbf{X}^t$ auto-regressively conditioned on the corrupted source sequence $\hat{\mathbf{X}}^t$ as
\begin{equation}
    \mathcal{L}_{mle}^{t} = \sum_{n =1}^{N^t} \log p (\mathbf{y}_n^t|\mathbf{y}_{<n}^t,\hat{\mathbf{X}}^t),
\end{equation}
% where the target sequence $\mathbf{Y} \in \mathcal{D}^\text{T}$.

% \begin{figure}[t]
% 	\centering
% 	\includegraphics[width=7.5cm]{codebook.png}
% 	\caption{The joint pre-training approach. The shared quantization module over encoder representations produces cross-modality quantized latent speech/text units. Hidden states and latent units are mixed up and used as the inputs of the cross-attention module in the decoder. The model learns to share discrete tokens across modalities, creating bridges between speech and text.
% 	}\label{codebook}
% \end{figure}

\paragraph{Joint Pre-Training}

The above pre-training methods can only leverage speech or text data to model acoustic or language information individually. To build a cross-modality mapping between speech and text, which is essential for tasks such as ASR and TTS, we propose a cross-modal vector quantization method to learn representations capturing the modality-invariant information.

Specifically, we utilize vector quantized embeddings as a bridge to align the speech representation and text representation through a shared codebook, as shown in Figure \ref{architecture}(b).
Inspired by VQ-VAE \cite{oord2017neural} and SemFace \cite{ren2021semface}, we first use the quantizer to convert these continuous speech/text representations $\mathbf{u}_i$ from the output of the encoder into discrete representations $\mathbf{c}_i$ from a fixed-size codebook $\mathbf{C}^K$, which contains $K$ learnable embeddings.
Then, the nearest neighbor search is performed between the encoder output and the embedding of each latent code via the $L_2$ distance as
\begin{equation}
    \mathbf{c}_i = \arg\min_{j\in[K]} \|\mathbf{u}_i-\mathbf{c}_j\|_2,
\end{equation}
where $\mathbf{c}_j$ is the $j$-th quantized vector in the codebook.
Note that we do the same operation for the output of the speech and text encoders with a shared codebook.

Then, we randomly replace a proportion (10\%) of the contextual representations with quantized latent representations in the corresponding time steps and calculate the cross-attention upon the mixed representations, which can explicitly guide the quantizer to utilize the cross-modal information.
The diversity loss is used to encourage sharing more codes by maximizing the entropy of the averaged Softmax distribution as
\begin{equation}
    \mathcal{L}_d= \frac{1}{K} \sum_{k=1}^{K} p_k \log p_k,
\end{equation}
where $p_k$ is the averaged probability of choosing the $k$-th code in the codebook. 

% More importantly, we  propose use gemerative adversarial network (GAN) \cite{goodfellow2014generative} to map speech and text into the same semantic space and try to align them.
% The discriminator $f_D$ takes mixed representation as input and outputs a confidence score between 0 and 1 to differentiate the representation mapped from speech encoder and text decoder during training.
% The encoder module $f_G$ acts as the generator whose goal is to make $f_G$ able to provide indistinguishable speech representation and text representation, to fool the discriminator.
% Formally, the adversarial training objective $L_{adv}$ can be written as:
% \begin{equation}
% \begin{aligned}
%     \mathcal{L}_{adv} = &\min_G \max_D \{ \mathbb{E} (log f_D(x)) \\ 
%     &+ \mathbb{E} (log(1-f_D(f_G(x)))) \} 
% \end{aligned}
% \end{equation}

The final pre-training loss with unlabeled speech and text data can be formulated as
\begin{equation}
\label{overall_loss}
    %\mathcal{L} =  \mathcal{L}_{mlm}^{s} + \mathcal{L}_{1}^{s} + \mathcal{L}_{2}^{s} + \mathcal{L}_{mle}^{t} + \mathcal{L}_{d} + \mathcal{L}_{adv}
    % \mathcal{L} = \mathcal{L}_{1}^{s} + \mathcal{L}_{2}^{s} +  \mathcal{L}_{bce}^{s} + \mathcal{L}_{mle}^{t} + \mathcal{L}_{d}.
    \mathcal{L} = \mathcal{L}_{mlm}^{s} + \mathcal{L}_{1}^{s} +  \mathcal{L}_{bce}^{s} + \mathcal{L}_{mle}^{t} + \gamma \mathcal{L}_{d}.
\end{equation}
where $\gamma$ is set to $0.1$ during pre-training.

\begin{table*}[!h]
\begin{center}
%\small
\small
\begin{tabular}{lcccccc}
    \toprule
     Model & LM & dev-clean &dev-other&test-clean&test-other\\ % \cline{3-4} \cline{5-6}
    % &&clean&other&clean&other \\
    \midrule\midrule
    % \multicolumn{7}{c}{\textbf{\textit{10-hour labeled}}} \\
    % \midrule
    % wav2vec 2.0 \textsc{Base} \cite{baevski2020wav2vec} & LS-960 & - & 10.9 & 17.4 & 11.1 & 17.6 \\
    % HuBERT \textsc{Base} \cite{hsu2021hubert} \dagger & LS-960 & - & 9.6 & 16.3 & 9.4 & 16.7 \\
    % Baseline & LS-960 & - &  &  & &  \\
    % SpeechT5 \textsc{Base} & LS-960 + LS-LM & - &  &  &  &  \\
    % \hline
    % DeCoAR 2.0 \cite{ling2020decoar} & LS-960 & 4-gram & - & - & 5.4 & 13.3 \\
    % wav2vec 2.0 \textsc{Base} \cite{baevski2020wav2vec} & LS-960 & 4-gram & 3.8 & 9.1 & 4.3 & 9.5 \\
    % % wav2vec 2.0 \textsc{Large} \cite{baevski2020wav2vec}  & LL-60k & - & 3.3 & 6.5 & 3.1 & 6.3 \\
    % HuBERT \textsc{Base} \cite{hsu2021hubert} & LS-960 & 4-gram & 3.9 & 9 & 4.3 & 9.4 \\
    % wav2vec 2.0 \textsc{Base} \cite{baevski2020wav2vec} & LS-960 & Transf. & 2.9 & 7.4 & 3.2 & 7.8 \\
    % wav2vec 2.0 \textsc{Large} \cite{baevski2020wav2vec}  & LL-60k & Transf. & 2.4 & 4.8 & 2.6 & 4.9 \\
    % HuBERT \textsc{Large} \cite{hsu2021hubert} & LL-60k & Transf. & 2.2 & 4.3 & 2.4 & 4.6 \\
    % Baseline & LS-960 & Transf. &  &  &  &  \\
    % SpeechT5 \textsc{Base} & LS-960 + LS-LM & Transf. &  &  &  &  \\
    % % SpeechT5 \textsc{Base} & LL-60k & - &  &  &  &  \\
    % SpeechT5 \textsc{Base} & LL-60k + LS-LM & Transf. &  &  &  & \\
    % % SpeechT5 \textsc{Large} & LL-60k & - &  &  &  & \\
    % SpeechT5 \textsc{Large} & LL-60k + LS-LM & Transf. &  &  &  & \\
    % \midrule\midrule
    % \multicolumn{7}{c}{\textbf{\textit{100-hour labeled}}} \\
    % \midrule
    wav2vec 2.0 \textsc{Base} \cite{baevski2020wav2vec} & - & 6.1 & 13.5 & 6.1 & 13.3 \\
    HuBERT \textsc{Base} \cite{hsu2021hubert} $\dagger$ & - & 5.5 & 13.1 & 5.8 & 13.3 \\
    Baseline (w/o CTC) & - & 5.8 & 12.3 & 6.2 & 12.3 \\
    Baseline & - & 4.9 & 11.7 & 5.0 & 11.9 \\
    SpeechT5 (w/o CTC) & - & 5.4 & 10.7 & 5.8 & 10.7 \\
    SpeechT5 & - & \textbf{4.3} & \textbf{10.3} & \textbf{4.4} & \textbf{10.4} \\
    \hline
    % DeCoAR 2.0 \cite{ling2020decoar}& 4-gram & - & - & 5.0 & 12.1 \\
    DiscreteBERT \cite{baevski2019effectiveness} & 4-gram & 4.0 & 10.9 & 4.5 & 12.1 \\
    wav2vec 2.0 \textsc{Base} \cite{baevski2020wav2vec}  & 4-gram & 2.7 & 7.9 & 3.4 & 8.0 \\
    HuBERT \textsc{Base} \cite{hsu2021hubert} & 4-gram & 2.7 & 7.8 & 3.4 & 8.1 \\
    wav2vec 2.0 \textsc{Base} \cite{baevski2020wav2vec} & Transf. & 2.2 & 6.3 & 2.6 & 6.3 \\
    % wav2vec 2.0 \textsc{Large} \cite{baevski2020wav2vec}  & LL-60k & - & 3.3 & 6.5 & 3.1 & 6.3 \\
    % wav2vec 2.0 \textsc{Large} \cite{baevski2020wav2vec}  & LL-60k & Transf. & 1.9 & 4.0 & 2.0 & 4.0 \\
    % HuBERT \textsc{Large} \cite{hsu2021hubert} & LL-60k & Transf. & 1.8 & 3.7 & 2.1 & 3.9 \\
    Baseline & Transf. & 2.3 & 6.3 & 2.5 & 6.3 \\
    SpeechT5 & Transf. & \textbf{2.1} & \textbf{5.5} & \textbf{2.4} & \textbf{5.8} \\
    % SpeechT5 \textsc{Base} & LL-60k & - &  &  &  &  \\
    % SpeechT5 \textsc{Base} & LL-60k + LS-LM & Transf. &  &  &  & \\
    % SpeechT5 \textsc{Large} & LL-60k & - &  &  &  & \\
    % SpeechT5 \textsc{Large} & LL-60k + LS-LM & Transf. &  &  &  & \\
    \bottomrule
\end{tabular}
\end{center}
\caption{\label{exp_asr} Results of ASR (speech to text) on the LibriSpeech dev and test sets when training on the 100 hours subset of LibriSpeech. 
% Baselines have the same model architecture as SpeechT5, but the encoder is initialized by the HuBERT \textsc{Base} model. 
% w/o CTC means that the joint CTC/attention decoding is not applied. 
$\dagger$ indicates that results are not reported in the corresponding paper and evaluated by ourselves.}
\end{table*}

\subsection{Fine-Tuning}
\label{sec_finetuning}

After pre-training, we fine-tune the encoder-decoder backbone via the loss of the downstream task.
The goal is to measure the learning abilities of SpeechT5, and we study the performance on a diverse set of downstream tasks such as ASR, TTS, ST, VC, SE, and SID.
% To specify which task the model should perform, we add a task-specific prefix to the original input sequence before feeding it to the model.
All of the spoken language processing tasks that we consider can be learned by concatenating the outputs of the encoder-decoder backbone and corresponding pre-net and post-net. 
Taking ASR as an example, the final model consists of the speech-encoder pre-net, encoder-decoder, text-decoder pre-net, and text-decoder post-net, which are initialized by SpeechT5 and fine-tuned via the cross-entropy loss on the corresponding training data. 
%For text to speech tasks such as TTS, the text pre-net and the speech post-net are introduced. For speech to speech tasks such as VC, the speech pre-net and post-net are used. For speech to class tasks such as SID, the architecture with the same components as speech to text tasks is considered, 
The baseline systems have the same architecture as SpeechT5, but the weights of the baseline encoder are initialized by the HuBERT \textsc{Base} model \cite{hsu2021hubert} if the input data of the downstream tasks is speech. 
It allows raw waveform as the model input and can provide a strong baseline.

\section{Experiments}

\subsection{Pre-Training Setup}

All models are implemented in Fairseq{\footnote[4]{https://github.com/pytorch/fairseq}} \cite{ott2019fairseq}.
% {\footnote[2]{https://github.com/pytorch/fairseq}}
The encoder-decoder backbone contains 12 Transformer encoder blocks and 6 Transformer decoder blocks, where the model dimension is 768, the inner dimension (FFN) is 3,072, and the number of attention heads is 12.
%, whose encoder is the same as the HuBERT \textsc{Base}.
The above encoder setting is the same as that in wav2vec 2.0 \textsc{Base} and HuBERT \textsc{Base}.
%Speech-encoder pre-net is two 1-dimensional convolutional layers with strides [2,2], kernel size [5,5] and channel size [1024,1536], where each layer is followed by a gated linear unit \cite{yann2017glu}. 
The speech-encoder pre-net contains 7 blocks of temporal convolutions, each of which is composed of 512 channels with strides $(5,2,2,2,2,2,2)$ and kernel sizes $(10,3,3,3,3,2,2)$.
For the speech-decoder pre-net and post-net, we use the same setting as the pre-net and post-net in \citet{shen2018tacotron2} except that the number of channels of the post-net is 256.
For text-encoder/decoder pre/post-net, a shared embedding layer with dimension 768 is used.
% For the vector quantization, we use 2%$G=2$ 
% codebooks with $V=100$ entries for the shared codebook module, resulting in a theoretical maximum of 10k codewords.
For the vector quantization, we use two codebooks with 100 entries for the shared codebook module, resulting in a theoretical maximum of $K=10^4$ code entries.

%The speech feature is a sequence of 80-dim log Mel-filterbank with 64 millisecond (ms) window, and 16 ms frame shift. It is normalized with utterance-level mean and variance when used as input data.
For speech pre-training, we use the full 960 hours of LibriSpeech audio \cite{panayotov2015librispeech}. %or 60K hours of Libri-light dataset \cite{kahn2020libri}.
For text pre-training, we use the normalized language model training text of LibriSpeech as unlabeled data, which contains 400M sentences.{\footnote[5]{https://www.openslr.org/11}}
We optimize the model with Adam \cite{kingma2014adam} by warming up the learning rate for the first 8\% of updates to a peak of $2 \times 10 ^{-4}$, which is linear decayed for the following updates.
% We set the weight $\lambda = 0.1$ for the diversity loss in Equation \ref{overall_loss}.
We pre-train the proposed SpeechT5 model on 32 V100 GPUs with a batch size of around 90s samples per GPU for speech and 12k tokens per GPU for text and set the update frequency to 2 for 500k steps.

%For unsupervised speech representation learning, we use the full 960 hours of LibriSpeech audio \cite{panayotov2015librispeech}, which is derived from the LibriVox project that contains English recordings of copyright-free audiobooks by volunteers from the Internet.
%For unsupervised text representation learning, we use the normalized language model training text of LibriSpeech as unlabeled data, which contains 400M sentences.{\footnote[3]{\url{https://www.openslr.org/11/}}}

\begin{table*}[!htp]
\begin{center}
\small
\begin{tabular}{@{\extracolsep{2pt}}lcccc@{}}
    \toprule
     \multirow{2}{*}{Model} & \multicolumn{2}{c}{WER} & \multicolumn{2}{c}{MCD} \\
        \cline{2-3} \cline{4-5}
        & bdl to slt & clb to slt & bdl to slt & clb to slt \\
    \midrule
    \midrule
    VTN w/ ASR  \cite{huang2021pretraining} &11.1\% & 10.9\% & 6.50 & 6.11 \\
    VTN w/ TTS  \cite{huang2021pretraining} & \textbf{7.6\%} & 9.1\% & 6.33 & 6.02 \\ % Report: The CER and WER for the ground-truth test set of slt were 0.9% and 3.8%, respectively.
    Many-to-many VTN \cite{Kameoka2021} & - & - & 6.13 & 5.97 \\
    \midrule
    Baseline & 21.5\% & 10.8\% & 6.26 & 6.16 \\ % ASR Model: WER: 3.01% CER: 0.54% on the vocoder-generated test set of slt.
    SpeechT5 & 7.8\% & \textbf{6.4\%} & \textbf{5.93} & \textbf{5.87} \\
    \bottomrule
\end{tabular}
\end{center}
\caption{\label{exp_vc} Results of VC (speech to speech) on the CMU Arctic. The bdl, clb, and slt denote three speakers.}
%  VTN \cite{huang2021pretraining} is the state-of-the-art voice Transformer network that is fine-tuned from the pretrained ASR or TTS model.
\end{table*}

\subsection{Evaluation on ASR}

%In supervised fine-tuning, we use the commonly adopted dataset and evaluation metric for each task.
We fine-tune the ASR model with the LibriSpeech 100/960 hours data and train the language model (LM) with the LibriSpeech LM text data, which is used for shallow fusion \cite{gulcehre2015using} during the ASR inference.
% Besides the cross-entropy loss at the decoder side, we add an extra linear layer to calculate the connectionist temporal classification (CTC) loss at the top of the encoder \cite{shinji2017hybrid} to enable the joint CTC and decoder inference \cite{hori-etal-2017-joint} to boost the performance.
Besides the cross-entropy loss for the decoder, we add an extra linear layer to calculate the connectionist temporal classification (CTC) loss on the top of the encoder \cite{shinji2017hybrid}, so that we can apply the joint CTC/attention decoding \cite{hori-etal-2017-joint} to boost the performance.
%We evaluate the results of ASR using word error rate (WER).
We measure the performance of ASR by the word error rate (WER).
%The word error rate (WER) is evaluated on the standard Librispeech dev-other/clean and test-clean/other sets.
The implementation details can be found in Appendix \ref{appendix_asr}.

The results of ASR on the 100 hours set of LibriSpeech are reported in Table \ref{exp_asr}. 
%The WER is evaluated on the standard Librispech dev-other/clean and test-clean/other sets.
We compare with several state-of-the-art self-supervised approaches, including DiscreteBERT \cite{baevski2019effectiveness}, wav2vec 2.0 \cite{baevski2020wav2vec}, and HuBERT \cite{hsu2021hubert}.
% The baseline system has the same model structure as SpeechT5, while the weights of the encoder are initialized by the HuBERT \textsc{Base} model.
Without LM fusion, the baseline outperforms wav2vec 2.0 \textsc{Base} and HuBERT \textsc{Base} with the help of the joint CTC/attention decoding, which shows the importance of the decoder.
The proposed SpeechT5 model achieves significant improvements on all settings compared to wav2vec 2.0 \textsc{Base}, HuBERT \textsc{Base} and our strong baselines, demonstrating the superiority of the proposed pre-training method.
Furthermore, when decoding with LM fusion, SpeechT5 obtains the lower WERs than wav2vec 2.0 \textsc{Base} on all sets and achieves the state-of-the-art performance.
Due to space constraints, the results of 960h fine-tuning experiments are reported in Appendix \ref{sec:appendix_960h}.

\subsection{Evaluation on TTS}

%For the TTS task, the SpeechT5 architecture consists of the text-encoder pre-net, encoder-decoder, speech-decoder pre-net, and speech-decoder post-net, which is fine-tuned via the $L_1$ loss and the BCE loss as mentioned in the speech reconstruction of the speech representation learning.

We fine-tune the pre-trained model on the 460-hours LibriTTS clean sets \cite{zen2019libritts} with the $L_1$ loss, $\mathcal{L}_{bce}^{s}$ loss, and attention loss \cite{tachibana2018efficiently}.
We utilize the HiFi-GAN \cite{kong2020hifi} vocoder to convert the log Mel-filterbank to the raw waveform.
We evaluate the Naturalness with the open-source NISQA-TTS \cite{Mittag2020}, the mean option score (MOS), and the comparison mean option score (CMOS) by native speakers on the randomly selected 200 sentences with various lengths (no overlapping with training data) generated by different models, in which case we keep the text content consistent. 
More details can be found in Appendix \ref{appendix_tts}.

\begin{table}[!h]
\begin{center}
\small
\begin{tabular}{l|ccc}
    \toprule
    % Model & WER & Naturalness & MOS & CMOS \\
    Model & Naturalness & MOS & CMOS \\
    \midrule
    % Ground Truth  & 4.1\% & & 3.86 $\pm$ 0.03 & - \\
    Ground Truth  & -\hspace{3ex} & 3.87 $\pm$ 0.04 & - \\
    %Baseline & 1.72 & 3.45 $\pm$ 0.04 -\\
    %SpeechT5 \textsc{Base} & \textbf{1.49} & \textbf{3.65} $\pm$ 0.04 & - \\
    % Baseline & 4.1\% & 2.76 & $\pm$ 0.04 & 0\\ % baseline5
    Baseline & 2.76 & 3.56 $\pm$ 0.05 & 0 \\ % baseline5
    % SpeechT5 \textsc{Base} & 4.1\% & 2.91 &  $\pm$ 0.04 &  \\ % finetune14 speecht5 w/o hubert
    SpeechT5 & \textbf{2.91} &  \textbf{3.65} $\pm$ 0.04 &  +\textbf{0.290} \\ % finetune14 speecht5 w/o hubert
    \bottomrule
\end{tabular}
\end{center}
\caption{\label{exp_tts} Results of TTS (text to speech) on the LibriTTS.}
%The naturalness, MOS, and CMOS are evaluated on 200 selected sentences.}
\end{table}

Table \ref{exp_tts} shows the experimental results of TTS. The proposed SpeechT5 trained without $\mathcal{L}_{mlm}^{s}$ is considered because the bidirectional masked prediction loss is proposed to help the encoder learn to encode the speech signal, and this variant achieves superior Naturalness, as shown in Table \ref{tab:tts_as} (in Appendix \ref{sec:appendix_tts_wo_hubert}).
% In Table \ref{exp_tts}, the baseline and the proposed SpeechT5 maintain the content of speech and obtain 4.1\% WER equal to that of the ground truth speech.
The proposed SpeechT5 model behaves better than baseline and achieves a performance of 2.91 Naturalness and 3.65 MOS.
Furthermore, our proposed SpeechT5 obtains a gain of +0.29 in CMOS with respect to the baseline model, which suggests the proposed pre-training method significantly improves the speech generation quality.

\subsection{Evaluation on ST}
We evaluate the ST task on the MUST-C dataset \cite{di-gangi-etal-2019-must}, including English-German (EN-DE) and English-French (EN-FR) translation tasks.
We use the default training setting of speech translation in Fairseq ST \cite{wang2020fairseq}, and we also average the last 10 checkpoints and use a beam size of 5 for decoding.
Translation results are evaluated with case-sensitive BLEU \cite{papineni2002bleu}. Details about the dataset and fine-tune setting are introduced in Appendix \ref{appendix_st}.

\begin{table}[!h]
\begin{center}
\small
\begin{tabular}{l|cc}
    \toprule
    Model & EN-DE & EN-FR \\
    \midrule
    Fairseq ST \cite{wang2020fairseq}  & 22.70 & 32.90 \\
    ESPnet ST \cite{inaguma2020espnet}  & 22.91 & 32.69 \\
    Adapter Tuning \cite{le-etal-2021-lightweight}  & 24.63 & 34.98 \\
    \midrule
    Baseline & 23.43 & 33.76  \\
    SpeechT5 (w/o initializing decoder) & 24.44 &  34.53 \\
    SpeechT5 & \textbf{25.18} &  \textbf{35.30} \\
    \bottomrule
\end{tabular}
\end{center}
\caption{\label{exp_st} Results of ST (speech to text) on the MUST-C EN-DE and EN-FR.}
\end{table}

% \paragraph{Dataset and Evaluation Metrics}
% We evaluate our ST task on MUST-C dataset \cite{di-gangi-etal-2019-must}, including English-German (EN-DE) and English-French (EN-FR) translation tasks.  The EN-DE/EN-FR language pair consists of 408/492 hours of speech data aligned with 234K/280K translated sentences.
% We report the results on EN-DE and EN-FR tst-COMMON set (2641 and 2632 utterances).
% Translation results are computed with case-sensitive BLEU \cite{papineni2002bleu}.

% \paragraph{Fine-Tuning Details}
% We use waveform as speech inputs for speech translation.
% The training setting is the same as that in S2T model in Fairseq.
% We use 8K unigram vocabulary for EN/DE and EN-FR, respectively.
% Following Fairseq ST \cite{wang2020fairseq}, we average the last 10 checkpoints and use a beam size of 5 for decoding.

%\paragraph{Main Results}
We list the BLEU scores of ST in Table \ref{exp_st}.
% The baseline is an encode-decoder model whose encoder is initialized by the HuBERT \textsc{Base} model.
The result of SpeechT5 without initializing the decoder is also reported since we do not pre-train the decoder with German or French data, and it outperforms the strong baseline whose encoder is initialized by HuBERT encoder.
The proposed SpeechT5 further beats the SpeechT5 without initializing the decoder, and achieves a significant improvement of 1.75 and 1.54 BLEU scores than baseline in EN-DE and EN-FR tasks, respectively, which demonstrates the effectiveness and superiority of our method.
%Although the data used for pre-training do not contain German and French data, the proposed SpeechT5 model achieves a significant improvement of 1.75 and 1.54 BLEU scores in the EN-DE and EN-FR tasks, respectively.
Besides, our SpeechT5 model outperforms existing models such as Fairseq ST \cite{wang2020fairseq}, ESPnet ST \cite{inaguma2020espnet}, and Adapter Tuning \cite{le-etal-2021-lightweight} that employs adapter modules to be further specialized in each language pair from different pre-trained models.

\subsection{Evaluation on VC}
% For fine-tuning experiments of 100 hours set, we set the $\alpha$ to 0.5 and $\beta$ to 1.0.

VC aims to convert a speaker-dependent source speech waveform into a different one while preserving linguistic information of the source speech waveform.
%For the VC task, the speech pre-net and post-net are used with the encoder-decoder as the SpeechT5 architecture of the speech to speech task, which allows raw audio as inputs and produces log Mel-filterbank.
We follow the many-to-many setting and utilize speech recordings of four speakers in the CMU Arctic \cite{kominek2004cmu}, including clb, bdl, slt, and rms.
For the waveform synthesis, we use the Parallel WaveGAN \cite{Yamamoto2020}, a non-autoregressive variant of the WaveNet vocoder. 
We employ the average of MCD (Mel-Cepstral Distortion) and WER as the metrics for the VC task.
% The average of MCD (Mel-Cepstral Distortion) token along the DTW (dynamic time warping) path between the output and ground-truth mel-cepstra serves as the evaluation metric of VC. 
% Also, we use WER to evaluate the generated voice with the public ASR model HuBERT \textsc{Large}{\footnote[5]{https://huggingface.co/facebook/hubert-xlarge-ls960-ft}}.
More details about the dataset and fine-tune setting are given in Appendix \ref{appendix_vc}.

We show the results of VC in Table \ref{exp_vc}, where we list the conversion from speaker bdl to slt and clb to slt as used in the voice Transformer network (VTN) \cite{huang2021pretraining}.
The experimental results demonstrate that the proposed SpeechT5 model achieves a significant gain than the strong baseline model.
The proposed SpeechT5 model also outperforms the state-of-the-art VTN variants in terms of MCD, including VTN fine-tuned from ASR or TTS \cite{huang2021pretraining} and many-to-many VTN \cite{Kameoka2021}.
% , where VTN  is a Transformer-based seq2seq VC model using pretrained ASR or TTS model parameters.

\subsection{Evaluation on SE}

SE is the task of removing background noise from a degraded speech signal and improving the intelligibility and the perceived quality of the signal.
We use the WSJ0 Hipster Ambient Mixtures (WHAM!) dataset \cite{Wichern2019} and conduct the 16 kHz max enhance-single task that recovers the signal from a mixture of only the first WSJ0 speaker and noise.
We utilize HiFi-GAN to transform the log Mel-filterbank to the raw waveform.
Since the input and output lengths are probably different in the encoder-decoder model, we can not evaluate it by PESQ \cite{Rix2001} and ESTOI \cite{Jensen2016}, so we evaluate the negative impact on the ASR performance by WER.
The implementation details of SE are in Appendix \ref{appendix_se}.

\begin{table}[!h]
\begin{center}
\small
\begin{tabular}{l|cc}
    \toprule
    Model & WER  \\
    \midrule
    Ground Truth Speech & 3.2\% \\
    Noisy Speech \cite{Wichern2019} & 76.1\% \\
    \midrule
    % MetricGAN+ \cite{fu2021metricgan+} & 78.6\% \\
    % NSNet2 \cite{braun2020data} & 45.8\% \\
    % Espnet SE \cite{li2021espnet} & 41.7\% \\
    % \midrule
    Baseline & 10.9\% \\
    SpeechT5 & \textbf{8.9\%}  \\
    \bottomrule
\end{tabular}
\end{center}
\caption{\label{tab:exp_se} Results of SE (speech to speech) on the WHAM!.}
\end{table}

As shown in Table \ref{tab:exp_se}, our strong baseline model recovers contents from the noisy speech, achieving 10.9\% WER from 76.1\% WER.
Moreover, the proposed SpeechT5 model gets a relative 9\% WER reduction compared to the strong baseline model.
%baseline and achieves a superior WER, which is close to the ground truth.
% compared to NSNet2 and Espnet SE, probably due to the mismatch of the training dataset in noise intensity between the compared methods and the proposed SpeechT5 as summarized in Table \ref{objective_noise_results}. 
The results suggest that although the noisy speech with WHAM! is challenging as summarized in Table \ref{objective_noise_results} (in Appendix \ref{appendix_se}), the proposed encoder-decoder framework can effectively suppress the noise and recover the content.

\subsection{Evaluation on SID}

%SID is the process of classifying the identity of an unknown voice among a set of speakers based on the speaker's known utterances. We utilized the SpeechT5 architecture as the speech to text task, where the model predicts the probability that a single text token belongs to a target class.

We convert SID, a multi-class classification task of classifying each utterance for its speaker identity, to a speech to text task by sequence to sequence model. 
% Specifically, the Transformer decoder and encoder take the \texttt{<sos>} token and $\mathbf{H}$ as the input, respectively. 
Compared to the ASR task, the text embedding table is replaced by a speaker embedding table, and the decoder predicts speaker identifies at the first step.
We adopt the VoxCeleb1 dataset \cite{nagrani2017voxceleb}, which contains over 100,000 speech records uttered by 1,251 celebrities extracted from videos uploaded to YouTube.
The top-1 speaker classification accuracy (ACC) is used as the evaluation metric of SID. Refer to Appendix \ref{appendix_sid} for more details about the dataset and fine-tuning.

\begin{table}[!h]
\begin{center}
\small
% \begin{tabular}{l|l|c}
\begin{tabular}{l|c}
    \toprule
    % Framework & Model & ACC \\
    Model & ACC \\
    \midrule
    % \multirow{3}{*}{SUPERB \cite{yang2021superb}}& wav2vec 2.0 Base \cite{baevski2020wav2vec} & 75.18\% \\
    %   & HuBERT Base \cite{hsu2021hubert} & 81.42\% \\
    %   & HuBERT Large \cite{hsu2021hubert} & 90.33\% \\
    SUPERB \cite{yang2021superb} \\
    \hspace{2ex}wav2vec 2.0 \textsc{Base} \cite{baevski2020wav2vec} & 75.18\% \\
    \hspace{2ex}HuBERT \textsc{Base} \cite{hsu2021hubert} & 81.42\% \\
    \hspace{2ex}HuBERT \textsc{Large} \cite{hsu2021hubert} & 90.33\% \\
    \midrule
    % \multirow{2}{*}{SpeechNet \cite{chen2021speechnet} } & Single   Task & 86.00\% \\
    %   & Multi-Task   with TTS & 87.90\% \\
    SpeechNet \cite{chen2021speechnet} \\
    \hspace{2ex}Single Task & 86.00\% \\
    \hspace{2ex}Multi-Task with TTS & 87.90\% \\
    % \midrule
    % \multirow{2}{*}{Ours (FBANK)}  & Baseline & 88.90\% \\
    %   & SpeechT5 & \textbf{90.97}\% \\
    \midrule
    % CNN-LDE \cite{Cai2018} & 89.9\% \\
    Thin ResNet-34 \cite{Chung2020} & 89.00\% \\
    \midrule
    Ours \\
    \hspace{2ex}Baseline & 91.92\% \\
    \hspace{2ex}SpeechT5 & \textbf{96.49}\% \\
    % \hspace{2ex}SpeechT5 \textsc{Large} &  \\
    \bottomrule
\end{tabular}
\end{center}
\caption{\label{exp_sid} Results of SID (speech to text) on the VoxCeleb1. The SUPERB fine-tuning freezes the encoder.}
%  SUPERB \cite{yang2021superb} is a leaderboard to benchmark the performance of a pre-trained model with minimal architecture changes and labeled data. SpeechNet \cite{chen2021speechnet} is a universal speech model with multi-task learning framework.
\end{table}

%\paragraph{Main Results}

% We compare our method to the scores of SUPERB \cite{yang2021superb} and SpeechNet \cite{chen2021speechnet}.
% SpeechNet is a universal model with multi-task learning framework, which attempts to improve the single task by utilizing multi-task learning.
% SUPERB is a benchmark across various speech tasks to evaluate pre-trained model. 
% In their leaderboard, wav2vec 2.0 \cite{baevski2020wav2vec} and HuBERT \cite{hsu2021hubert} are two state-of-the-art pre-trained models.
As shown in Table \ref{exp_sid}, our baseline is superior to existing Transformer-based methods such as SpeechNet \cite{chen2021speechnet} and pre-trained models from SUPERB \cite{yang2021superb}. Moreover, it outperforms ResNet-based architectures such as Thin ResNet-34 \cite{Chung2020}, indicating the superiority of the encoder-decoder architecture for the SID task. The SpeechT5 further improves the performance compared to baseline and achieves the state-of-the-art performance (i.e., 96.49\% accuracy), which demonstrates the effectiveness of the proposed pre-training technique.

\subsection{Ablation Study}

To better understand why the proposed SpeechT5 model is effective, we investigate the influence of the pre-training methods by removing each of them independently.
% in the pre-training process.

\begin{table}[!h]
\begin{center}
\small
\begin{tabular}{@{\extracolsep{5pt}}l|cccc@{}}
    \toprule
    % Model & VC & ASR & TTS & SID & SE \\
    \multirow{2}{*}{Model}  & \multicolumn{2}{c}{ASR} & \multirow{2}{*}{VC} & \multirow{2}{*}{SID}  \\ \cline{2-3}
    % &&&test-clean&test-other \\
    &clean&other&& \\
    \midrule
    % SpeechT5 & 6.14 & 11.93\% & 4.39\% & 95.35\% & \\ % 0.5 codebook
    SpeechT5& 4.4 & 10.7 & 5.93 & 96.49\%   \\ % 0.1 codebook
    \hspace{2ex}w/o Speech PT & - & - & 6.49 & 38.61\%  \\
    \hspace{2ex}w/o Text PT & 5.4 & 12.8 & 6.03 & 95.60\% \\
    \hspace{2ex}w/o Joint PT & 4.6 & 11.3 & 6.18 & 95.54\%  \\
    \hspace{2ex}w/o $\mathcal{L}_{mlm}^{s}$ & 7.6 & 22.4 & 6.29 & 90.91\% \\
    \bottomrule
\end{tabular}
\end{center}
\caption{\label{ablation} Ablation study for the SpeechT5 model.
Different variants of the SpeechT5 model, including the SpeechT5 model without speech pre-training (PT), text pre-training, joint pre-training method, or the bidirectional masked prediction loss, are evaluated on the ASR (test subsets with WER), VC (bdl to slt with MCD), and SID (test set with ACC) tasks.
}
\end{table}

As shown in Table \ref{ablation}, we can draw the following conclusions:
(1) The pre-training methods, including speech pre-training, text pre-training, and joint pre-training method, are important to SpeechT5 since without each of them, the performance of all tasks will degrade significantly; 
(2) Speech pre-training is more critical than text pre-training on these tasks that need to encode speech, and the ASR model fine-tuned from SpeechT5 without speech pre-training even can not converge;
(3) Without the joint pre-training method, the performance of the ASR model decreases, which demonstrates that the learned alignment from joint pre-training brings benefits for cross-modality tasks;
(4) The masked language model learning $\mathcal{L}_{mlm}^{s}$ of speech data is mainly responsible for extracting acoustic features and learning better speech representation, which is beneficial to ASR and SID tasks.

% \subsection{Effect of SpeechT5 Encoder}
% As a by-product of SpeechT5, the Transformer encoder of SpeechT5 can also be utilized as an independent pre-trained model for downstream tasks.
% We verify the performance of SpeechT5 encoder in SUPERB benchmark.  

% \paragraph{Speech pre-training vs. text pre-training}

% \paragraph{Effect of shared codebook}

% \paragraph{BART-style vs. T5-style noise}

% \paragraph{Quality of Sppech-T5 encoder}

% \subsubsection{Qualitative Analysis}

% To understand SpeechT5’s performance beyond automated metrics, we analyse its generations qualitatively.

\section{Related Work}

Large-scale pre-training models such as BERT \cite{devlin2018bert}, T5 \cite{raffel2019exploring}, wav2vec 2.0 \cite{baevski2020wav2vec}, and HuBERT \cite{hsu2021hubert} have drawn much attention in the NLP and speech communities, due to its strong capability of generalization and efficient usage of large-scale data \cite{devlin2018bert,liu2019roberta,yang2019xlnet,lewis2019bart,chen2021injecting,baevski2020wav2vec,lakhotia2021generative,kharitonov2021text,chen2021wavlm}.
%Recent pre-trained models in NLP, such as BERT \cite{devlin2018bert}, RoBERTa \cite{liu2019roberta}, XLNet \cite{yang2019xlnet} and BART \cite{lewis2019bart}, have achieved the state-of-the-art performance on language understanding and generation tasks.
%In spoken language processing, pre-trained speech models have also been applied to ASR \cite{hsu2021hubert}, TTS \cite{hayashi2019pre}, ST \cite{li2020multilingual}, VC \cite{huang2021pretraining}, and so on.
However, the research mentioned above effects gear towards single-modal learning,
% \footnote{***currently there are some multi-modal pre-trained models such as ViLT which operates on the text and image modalities. It is better to mention it and discuss the difference.} 
hence they can only be used in either text or speech modeling. 
Although some speech-language pre-training work \cite{chung2020splat,kim2021st,qian2021speech} attempts to improve spoken language understanding tasks,
%e.g., intent detection, dialog act classification, and spoken sentiment analysis, 
these methods only focus on an encoder with task-specific layers for different tasks and do not pre-train a decoder for generation tasks such as speech synthesis or text generation.
Besides, a series of research work begins to  investigate joint text and speech training \cite{han2021learning,ye2021end,tang-etal-2021-improving,zheng2021fused,tang2021general}, but they are mainly designed for  speech to text tasks. 

%Text pre-training, T5
The proposed SpeechT5 method is most related to T5 \cite{raffel2019exploring}. The core idea of the T5 model, a unified framework for a variety of text-based language problems, is to treat every text processing problem as a ``text-to-text'' problem.
%i.e., taking the text as the input and producing new text as the output.
%Unlike T5, SpeechT5 is a cross-modal encoder-decoder framework, whose input and output are speech or text through different pre/post networks.
%Besides, we propose a new joint speech-text pre-training method to leverage large-scale unlabeled text and speech dataset and align the textual and phonetic information.
SpeechT5 is also related to Speech Chain \cite{tjandra2020machine}, which leverages the ASR model and TTS model to build a closed-loop machine speech chain to train models on the concatenation of both labeled and unlabeled data, 
and SpeechNet \cite{chen2021speechnet}, which designs a universal modularized model to perform multiple speech processing tasks with multi-task learning.
The differences from the above models are that (1) SpeechT5 is a shared cross-modal encoder-decoder framework, whose input and output are speech or text through multiple pre/post-nets; (2) SpeechT5 attempts to pre-train and improve the universal model with large-scale unlabeled text and speech data.
%SpeechNet shows that it can simultaneously learn several common and important speech processing tasks.
%However, there are two big differences between SpeechNet and our SpeechT5. First, SpeechNet has a different encoder and decoder for different modalities  (e.g., speech and text), but SpeechT5 only uses one shared encoder-decoder model for all tasks. Second, SpeechNet aims to verify the multi-task learning in several speech tasks, but our SpeechT5 attempts to pre-train and improve the universal model with large-scale unlabeled text and speech data.

Another related work is SUPERB \cite{yang2021superb}, a benchmark to examine the capability of pre-trained models such as HuBERT \cite{hsu2021hubert}.
% collects various tasks with limited labeled data in speech communities to align with common research interests.
SUPERB focuses on investigating a simple framework to learn SUPERB tasks with a frozen and shared pre-trained encoder and lightweight prediction modules fine-tuned for each task.
In contrast, the goal of SpeechT5 is to learn all spoken language processing tasks by fine-tuning a unified-modal encoder-decoder model, which is pre-trained on unlabeled speech and text corpus.

%The most similar work to ours has been presented in

%\section{Conclusion and Future Work}
\section{Conclusion}

In this paper, we have proposed SpeechT5 as a pre-trained encoder-decoder model for various spoken language processing tasks.
We convert all spoken language processing tasks into a speech/text to speech/text format and propose a novel joint pre-training method to utilize cross-modal information by leveraging the unlabeled speech and text data.
The proposed unified encoder-decoder model can support generation tasks such as speech translation and voice conversion.
Massive experiments show that SpeechT5 significantly outperforms all baselines in several spoken language processing tasks.
In the future, we are going to pre-train the SpeechT5 with a larger model and more unlabeled data.
We are also interested in extending the proposed SpeechT5 framework to address multilingual spoken language processing tasks for future work.

% For future work, we plan to investigate more efficient pre-training methods,
% such as waveform learning representation via masked prediction like HuBERT \cite{hsu2021hubert}, aligning text token and phoneme explicitly as unsupervised ASR \cite{baevski2021unsupervised}.
% Besides, we will pre-train the SpeechT5 with a larger model and more unlabeled data, and fine-tune it on more spoken language processing tasks.
% We are also interested in extending the proposed SpeechT5 framework to address the multilingual spoken language processing problem.

\section*{Acknowledgments}

 We thank Yanqing Liu and Sheng Zhao for their help in TTS human evaluation. We also want to thank the anonymous reviewers for insightful comments and suggestions.

\bibliographystyle{acl_natbib}
% \bibliography{anthology,acl2021}
\bibliography{acl}

\clearpage

\appendix

\section{Comparisons of Text Mask Strategies}
\label{sec:appendix_mask}

% As shown in Table \ref{exp_asrmask} and \ref{tab:different_mask_vc_se_sid},
We compare the performance when using the BART \cite{lewis2019bart} or T5 \cite{raffel2019exploring} strategies for text masking on the ASR task, as reported in Table \ref{exp_asrmask}. The BART strategy achieves comparable or better performance than the T5 strategy under different inference settings. 
% As shown in \ref{tab:different_mask_vc_se_sid}, the BART strategy improves the performance in WER for the SE task and in ACC for the SID task. On the other hand, compared to T5 strategy, it brings little gain for the VC task, where the WER and MCD in the case of bdl to slt boosts, but the performance of clb to slt drops.

\section{Implementation Details}
\label{sec:appendix}

\subsection{ASR}
\label{appendix_asr}

\paragraph{Dataset}
% TODO: add more detail description of librispeech
% In supervised fine-tuning, we use the commonly adopted dataset and evaluation metric for each task.
%We train the ASR model by LibriSpeech 100 hours or 960h data,
We use the LibriSpeech corpus and fine-tune on two labeled data settings: 960 hours of transcribed Librispeech and the train-clean-100 subset comprising 100 hours (100 hours labeled).
We train the language model by the LibriSpeech language model (LM) text data, which is used for shallow fusion \cite{gulcehre2015using} during the ASR inference.

\paragraph{Fine-Tuning Details}

% Besides cross-entropy loss at the decoder side, we add an extra linear layer to calculate the CTC loss at the top of the encoder \cite{shinji2017hybrid}. 
We fine-tune the model with the CTC loss and the cross-entropy loss, where the loss weights are $0.5$  for both of them. 
We train on 8 V100 GPUs with a batch size of up to 256k audio samples per GPU. 
% The encoder is fixed for the first 13k steps.
The learning rate is warmed up for the first 10\% steps, hold as a constant for the following 40\% steps, and is decayed linearly for the rest steps.
Table \ref{exp_hpasr} summarizes the hyperparameters for ASR experiments of 100 hours and 960 hours sets.

\begin{table}[!h]
\begin{center}
\small
\begin{tabular}{l|cc}
    \toprule
    % Model & VC & ASR & TTS & SID & SE \\
    Hyperparameter & 100 hours & 960 hours \\
    \midrule
    % SpeechT5 & 6.14 & 11.93\% & 4.39\% & 95.35\% & \\ % 0.5 codebook
    updates & 80k & 320k \\ % 0.1 codebook
    learning rate & 6e-5 & 1.3e-4  \\
    time-step mask prob. & 0.075 & 0.05 \\
    channel mask prob & 0.008  & 0.0016 \\
    \bottomrule
\end{tabular}
\end{center}
\caption{\label{exp_hpasr} The setting of hyperparameters for ASR fine-tuning.}
\end{table}

\paragraph{Language Model and Decoding}
\label{lm_decoding}
We train a character-level LM for the ASR inference. 
The model has the same architecture as the Transformer LM in \citet{synnaeve2020endtoend}, which is used for decoding of wav2vec 2.0 \cite{baevski2020wav2vec} and HuBERT \cite{hsu2021hubert}. 
The LM contains 20 blocks of Transformer decoder with the model dimension of 1280, inner dimension of 6144, and 16 attention heads. 
% There are several differences between our LM and the LM in \cite{synnaeve2020endtoend}, such as the objective, batch size, and learning rate.
To investigate the difference of the performance between our LM and the LM in \citet{synnaeve2020endtoend}, we evaluate the word-level perplexities of these LMs on the LibriSpeech dev-clean/other sets as shown in Table \ref{exp_lm}. 
The Transformer LM used for SpeechT5 gets 56.5 perplexity on the dev-clean set and 59.3 perplexity on the dev-other set, which are higher than the perplexities of word Transformer LM in \citet{synnaeve2020endtoend}. 
It suggests that we may achieve better performance on the ASR task if the perplexities of our LM are similar to the LM in \citet{synnaeve2020endtoend}.

During decoding, the beam size is set to 30 for all experiments. 
We select the model with the highest accuracy on dev-other set for inference and apply the joint CTC/attention decoding \cite{hori-etal-2017-joint} to further improve the performance.
The model generates the output transcription by the beam search algorithm, which aims to maximize
\begin{equation}
   \alpha \log{P_{Dec}} + (1 - \alpha) \log{P_{CTC}} + \beta \log{P_{LM}}
\end{equation}
where $\alpha$ and $\beta$ are weights for the log probabilities, $P_{Dec}$, $P_{CTC}$, and $P_{LM}$ are the probabilities of the decoder, CTC, and LM, respectively.
We set $\alpha$ to 0.5 and $\beta$ to 1.0 for fine-tuning experiments of 100 hours set, and set $\alpha$ to 0.9 and $\beta$ to 0.7 for fine-tuning experiments of 960 hours set.

\begin{table}[t]
\begin{center}
\small
\begin{tabular}{l|cc}
    \toprule
    % \multirow{2}{*}{Model} &  \multirow{2}{*}{\makecell{Unlabeled Data}} & \multirow{2}{*}{LM} & \multicolumn{2}{c}{dev}& \multicolumn{2}{c}{test}\\ \cline{4-5} \cline{6-7}
    \multirow{2}{*}{Language Model} & \multicolumn{2}{c}{dev} \\ &clean & other \\
    \midrule
    Word 4-gram \cite{synnaeve2020endtoend}  & 148.0 & 136.6 \\
    Word Transf. \cite{synnaeve2020endtoend} & 48.2 & 50.2  \\
    Character Transf. & 56.5 & 59.3 \\
    \bottomrule
\end{tabular}
\end{center}
\caption{\label{exp_lm} Word-level perplexities of language models on dev-clean/other sets of LibriSpeech.}
\end{table}

\begin{table*}[!h]
\begin{center}
\small
%\small
\begin{tabular}{@{\extracolsep{2pt}}lccccccc@{}}
    \toprule
     \multirow{2}{*}{Mask Strategies} &  \multirow{2}{*}{\makecell{CTC}} & \multirow{2}{*}{\makecell{LM}} & \multicolumn{2}{c}{dev}& \multicolumn{2}{c}{test}\\ \cline{4-5} \cline{6-7}
    &&&clean&other&clean&other \\
    \midrule\midrule
    % wav2vec 2.0 \textsc{Base} \cite{baevski2020wav2vec} & N/A & - & 6.1 & 13.5 & 6.1 & 13.3 \\
    % HuBERT \textsc{Base} \cite{hsu2021hubert} & N/A & 4-gram & 2.7 & 7.8 & 3.4 & 8.0 \\
    % wav2vec 2.0 \textsc{Base} \cite{baevski2020wav2vec} & N/A & 4-gram & 2.7 & 7.9 & 3.4 & 8.0 \\
    % wav2vec 2.0 \textsc{Base} \cite{baevski2020wav2vec} & N/A & Transf. & 2.2 & 6.3 & 2.6 & 6.3 \\
    % \midrule
    % Bart Strategy & -& - & - & 5.5 & 11.1 & 6.0 & 11.1 \\
    % & -& \checkmark& - & 4.2 & 10.4 & 4.3 & 10.5 \\
    % & -& \checkmark & \checkmark & 2.3 & 5.7 & 2.4 & 6.0 \\
    BART \cite{lewis2019bart} & - & - & 5.4& 10.7 & 5.8 & 10.7 \\
    & \checkmark & - & 4.3 & 10.3 & 4.4 & 10.4 \\
    & \checkmark & \checkmark & 2.1 & 5.5 & 2.4 & 5.8 \\
    \midrule
    % T5 Strategy & - && -& 5.3 & 11.5 & 5.7 & 11.7\\
    % & -& \checkmark& - & 4.4 & 11.2 & 4.6 & 11.3 \\
    % & -& \checkmark & \checkmark & 2.3 & 6.0 & 2.5 & 6.3 \\
    T5 \cite{raffel2019exploring} &  - & - & 5.4 & 11.3 & 5.7 & 11.3 \\
    & \checkmark & - & 4.3 & 10.7 & 4.4 & 10.7 \\
    & \checkmark & \checkmark & 2.3 & 5.8 & 2.3 & 5.8 \\
    % \midrule
    % Bart Style & - & - & \checkmark & & & & \\
    % & -& \checkmark & \checkmark & & & & \\
    % & \checkmark& - & \checkmark && & & \\
    % & \checkmark& \checkmark & \checkmark && & & \\
    % \cdashline{1-8}
    % T5 Style & -& & \checkmark& & & & \\
    % & -& \checkmark & \checkmark & & & & \\
    % & \checkmark& - & \checkmark && & & \\
    % & \checkmark& \checkmark & \checkmark && & & \\
    \bottomrule
\end{tabular}
\end{center}
\caption{\label{exp_asrmask} Comparisons of mask strategies for the text pre-training under different inference settings. Models are pre-trained on the 960 hours speech data of LibriSpeech and 400M text sentences of LibriSpeech-LM corpus, and fine-tuned on the 100 hours labeled data of LibriSpeech. CTC and LM mean the Joint CTC/attention decoding \cite{hori-etal-2017-joint}, and language model fusion, respectively.}
\end{table*}

% \begin{table*}[!htp]
% \begin{center}
% \small
% \begin{tabular}{@{\extracolsep{2pt}}lcccccc@{}}
%     \toprule
%      \multirow{2}{*}{Mask Strategie} & \multicolumn{2}{c}{WER (VC)} & \multicolumn{2}{c}{MCD (VC)} & \multirow{2}{*}{WER (SE)} & \multirow{2}{*}{ACC (SID)} \\
%         \cline{2-3} \cline{4-5}
%         & bdl to slt & clb to slt & bdl to slt & clb to slt \\
%     \midrule
%     \midrule
%     BART \cite{lewis2019bart} & 7.8\% & 6.4\% & 5.93 & 5.87 & 8.7\% & 96.49\% \\
%     T5 \cite{raffel2019exploring} & 9.3\% & 5.9\% & 5.95 & 5.85 & 8.9\% & 95.69\% \\
%     \bottomrule
% \end{tabular}
% \end{center}
% \caption{\label{tab:different_mask_vc_se_sid} Comparisons of mask strategies for the text representation learning on the VC, SE, and SID tasks. Models are fine-tuned on the labeled corpus of the VC, SE, and SID tasks, respectively. The WER and MCD are evaluated for the VC task. The WER is evaluated for the SE task. The ACC is evaluated for the SID task.}
% \end{table*}

% \paragraph{Comparisons of Text Mask Strategies}
% We compare the ASR performance when using BART \cite{lewis2019bart} or T5 \cite{raffel2019exploring} strategies for text masking, as shown in table \ref{exp_asrmask}.
% Under different inference settings, BART strategy achieves comparable or better performance than T5 strategy.

\subsection{TTS}
\label{appendix_tts}

\paragraph{Dataset and Evaluation Metrics}
We use the 460-hours LibriTTS clean sets \cite{zen2019libritts}, a multispeaker corpus of read English speech from the audiobooks of the LibriVox project, as TTS training dataset.
We trim the waveform as ESPnet recipe \cite{watanabe2018espnet}.
The WER is evaluated by using the open-source ASR model wav2vec 2.0 CTC{\footnote[6]{https://huggingface.co/facebook/wav2vec2-base-960h}}.
The naturalness of synthetic speech is estimated by using the open-source TTS naturalness prediction model NISQA-TTS{\footnote[7]{https://github.com/gabrielmittag/NISQA}} \cite{Mittag2020}.

\paragraph{Fine-Tuning Details}
Besides the $L_1$ loss and BCE loss, we add an additional attention loss \cite{tachibana2018efficiently} to speed up model convergence.
% We apply the loss function as used in the fine-tuning of the VC task.
We train on 8 V100 GPUs in a speaker-independent manner by using the training data of the LibriTTS.
The model is updated for 120k steps with a learning rate of 0.0004, while each GPU processes up to 45,000 tokens for a batch. The learning rate is warmed up for the first 10k steps and decayed in an inverse square root manner for the rest steps.

\begin{table*}[!h]
\begin{center}
%\small
\small
\begin{tabular}{lcccccc}
    \toprule
     Model & LM & dev-clean &dev-other&test-clean&test-other\\ % \cline{3-4} \cline{5-6}
    % &&clean&other&clean&other \\
    \midrule\midrule
    % \multicolumn{7}{c}{\textbf{\textit{10-hour labeled}}} \\
    % \midrule
    % wav2vec 2.0 \textsc{Base} \cite{baevski2020wav2vec} & LS-960 & - & 10.9 & 17.4 & 11.1 & 17.6 \\
    % HuBERT \textsc{Base} \cite{hsu2021hubert} \dagger & LS-960 & - & 9.6 & 16.3 & 9.4 & 16.7 \\
    % Baseline & LS-960 & - &  &  & &  \\
    % SpeechT5 \textsc{Base} & LS-960 + LS-LM & - &  &  &  &  \\
    % \hline
    % DeCoAR 2.0 \cite{ling2020decoar} & LS-960 & 4-gram & - & - & 5.4 & 13.3 \\
    % wav2vec 2.0 \textsc{Base} \cite{baevski2020wav2vec} & LS-960 & 4-gram & 3.8 & 9.1 & 4.3 & 9.5 \\
    % % wav2vec 2.0 \textsc{Large} \cite{baevski2020wav2vec}  & LL-60k & - & 3.3 & 6.5 & 3.1 & 6.3 \\
    % HuBERT \textsc{Base} \cite{hsu2021hubert} & LS-960 & 4-gram & 3.9 & 9 & 4.3 & 9.4 \\
    % wav2vec 2.0 \textsc{Base} \cite{baevski2020wav2vec} & LS-960 & Transf. & 2.9 & 7.4 & 3.2 & 7.8 \\
    % wav2vec 2.0 \textsc{Large} \cite{baevski2020wav2vec}  & LL-60k & Transf. & 2.4 & 4.8 & 2.6 & 4.9 \\
    % HuBERT \textsc{Large} \cite{hsu2021hubert} & LL-60k & Transf. & 2.2 & 4.3 & 2.4 & 4.6 \\
    % Baseline & LS-960 & Transf. &  &  &  &  \\
    % SpeechT5 \textsc{Base} & LS-960 + LS-LM & Transf. &  &  &  &  \\
    % % SpeechT5 \textsc{Base} & LL-60k & - &  &  &  &  \\
    % SpeechT5 \textsc{Base} & LL-60k + LS-LM & Transf. &  &  &  & \\
    % % SpeechT5 \textsc{Large} & LL-60k & - &  &  &  & \\
    % SpeechT5 \textsc{Large} & LL-60k + LS-LM & Transf. &  &  &  & \\
    % \midrule\midrule
    % \multicolumn{7}{c}{\textbf{\textit{100-hour labeled}}} \\
    % \midrule
    wav2vec 2.0 \textsc{Base} \cite{baevski2020wav2vec} & - & 3.2 & 8.9 & 3.4 & 8.5 \\
    % HuBERT \textsc{Base} \cite{hsu2021hubert} \dagger & - &  &  & & \\
    Baseline (w/o CTC) & - & 3.1 & 7.8 & 3.1 & 7.6 \\
    Baseline & - & 2.8 & 7.6 & 2.8 & 7.4 \\
    SpeechT5 (w/o CTC) & - & 2.8 & 7.6 & 3.1 & 7.3 \\
    SpeechT5 & - & \textbf{2.5} & \textbf{7.4} & \textbf{2.7} & \textbf{7.1} \\
    \hline
    % DeCoAR 2.0 \cite{ling2020decoar} & LS-960 & 4-gram & - & - & & \\
    % DiscreteBERT \cite{baevski2019effectiveness} & LS-960 & 4-gram & & & & \\
    wav2vec 2.0 \textsc{Base} \cite{baevski2020wav2vec} & 4-gram & 2.0 & 5.9 & 2.6 & 6.1 \\
    % HuBERT \textsc{Base} \cite{hsu2021hubert} & LS-960 & 4-gram & & & & \\
    wav2vec 2.0 \textsc{Base} \cite{baevski2020wav2vec} & Transf. & 1.8 & 4.7 & 2.1 & 4.8 \\
    % wav2vec 2.0 \textsc{Large} \cite{baevski2020wav2vec}  & LL-60k & - & 3.3 & 6.5 & 3.1 & 6.3 \\
    % wav2vec 2.0 \textsc{Large} \cite{baevski2020wav2vec}  & LL-60k & Transf. & & & & \\
    % HuBERT \textsc{Large} \cite{hsu2021hubert} & LL-60k & Transf. & & & & \\
    Baseline & Transf. & 2.0 & 4.5 & 1.9 & 4.5 \\
    SpeechT5 & Transf. & \textbf{1.8} & \textbf{4.3} & \textbf{1.9} & \textbf{4.4} \\
    % SpeechT5 \textsc{Base} & LL-60k & - &  &  &  &  \\
    % SpeechT5 \textsc{Base} & LL-60k + LS-LM & Transf. &  &  &  & \\
    % SpeechT5 \textsc{Large} & LL-60k & - &  &  &  & \\
    % SpeechT5 \textsc{Large} & LL-60k + LS-LM & Transf. &  &  &  & \\
    \bottomrule
\end{tabular}
\end{center}
\caption{\label{exp_asr960} WER of ASR when training on the 960 hours labeled data of LibriSpeech. }
\end{table*}

\subsection{ST}
\label{appendix_st}

\paragraph{Dataset and Evaluation Metrics}
We evaluate the ST task on the MUST-C dataset \cite{di-gangi-etal-2019-must}, including English-German (EN-DE) and English-French (EN-FR) translation tasks.  The EN-DE/EN-FR language pair consists of 408/492 hours of speech data aligned with 234K/280K translated sentences.
We report the results on EN-DE and EN-FR tst-COMMON set (2641 and 2632 utterances).
Translation results are computed with case-sensitive BLEU \cite{papineni2002bleu}.

\paragraph{Fine-Tuning Details}
ST translates speech signals in a language to text in another language.
We use raw audio as speech inputs in our experiments.
The training setting is the same as that in S2T model in Fairseq.
We set training steps to 80K and warm-up steps to 10K.
Baseline and SpeechT5 models are trained with 8 GPUs via Adam optimizer.
We use 8K unigram vocabulary for both EN-DE and EN-FR.
Following Fairseq ST \cite{wang2020fairseq}, we average the last 10 checkpoints and use a beam size of 5 for decoding.

\subsection{VC}
\label{appendix_vc}
\paragraph{Dataset and Evaluation Metrics}
We consider the many-to-many setting for the CMU Arctic \cite{kominek2004cmu}, which contains speech recordings of four speakers, such as clb (female), bdl (male), slt (female), and rms (male), who read the same 1,132 phonetically balanced English utterances. Thus, there are twelve different combinations of source and target speakers. 
For each speaker, the first 932, the last 100, and the rest 100 sentences of the 1,132 sentences are used for training, test, and validation as \cite{huang2021pretraining}, respectively.
The average of MCD is estimated by using the DTW (dynamic time warping) path between the output and ground-truth Mel-cepstra. A smaller MCD indicates better performance.
The WER is evaluated by using the public ASR model HuBERT \textsc{Large}{\footnote[8]{https://huggingface.co/facebook/hubert-xlarge-ls960-ft}}, where the WER of the test set with this ASR model is comparable to that of VTN \cite{huang2021pretraining}.

\paragraph{Fine-Tuning Details}
Besides the $L_1$ loss and BCE loss, we add an additional attention loss \cite{tachibana2018efficiently} to speed up the model convergence. The model is trained on 8 V100 GPUs by the Adam optimizer with a batch size of 20000 tokens per GPU. We assign the learning rate based on the inverse square root with the maximum learning rate of $10^{-4}$ within 60k steps and apply 6k warm-up steps. 

\subsection{SE}
\label{appendix_se}

\begin{table}[t]
\begin{center}
\small
\begin{tabular}{l|c}
    \toprule
    Metric & WHAM! \\
    \midrule
    PESQ & 1.12 \\
    ESTOI & 0.48 \\
    WER (NSNet2 \cite{braun2020data}) & 45.8\% \\
    % WER (Espnet SE \cite{li2021espnet}) & 41.7\% \\
    % DNS-Challenge \cite{Reddy2021} & 2.19 & 0.59  \\
    \bottomrule
\end{tabular}
\end{center}
\caption{\label{objective_noise_results} Results of noisy speech utterances on the test set in terms of PEQS, ESTOI, and WER.}
\end{table}

\paragraph{Dataset and Evaluation Metrics}
We aim to recover the content of signals contaminated by various noises and reduce the negative impact on the performance of ASR systems.
The 16 kHz enhance-single task of the WHAM! dataset \cite{Wichern2019} is used as the SE dataset. It contains 20,000 training utterances, 5,000 validation utterances, and 3,000 test utterances, where the input waveform is a mixture of only the first WSJ0{\footnote[9]{https://catalog.ldc.upenn.edu/LDC93S6A}} speaker and noise. We trim the noisy segment without contents.
The WER is evaluated by using the open-source ASR model{\footnote[10]{https://doi.org/10.5281/zenodo.4243201}} because lengths of inputs and outputs are probably different in the encoder-decoder model.
% , and MetricGAN+{\footnote[9]{https://huggingface.co/speechbrain/metricgan-plus-voicebank}}, 
% where NSNet2, Espnet SE are from the 2020 Deep Noise Suppression (DNS) challenge \cite{Reddy2021}.
% , and MetricGAN+ is trained by the VoiceBank-DEMAND dataset \cite{Valentini-Botinhao2016}.
Since lengths of noisy speech utterances are the same as lengths of clean utterances, we measure the test set via speech quality (PESQ) \cite{Rix2001}, extended short-time objective intelligibility (ESTOI) \cite{Jensen2016}, and WER to quantify the difficulty of noisy speech, as shown in Table \ref{objective_noise_results}.
NSNet2
% and Espnet SE{\footnote[8]{https://doi.org/10.5281/zenodo.4923697}} are 
is the baseline model on the 2020 Deep Noise Suppression (DNS) challenge \cite{Reddy2021} and obtains WER of 45.8\%, probably due to the mismatch between the noise intensity of the WHAM! and DNS corpus.

\paragraph{Fine-Tuning Details} 
We employ the loss function as used in the fine-tuning of the VC task. The model is trained on 8 V100 GPUs by the Adam optimizer with a batch size of 16000 tokens per GPU. We assign the learning rate based on the inverse square root with the maximum learning rate $10^{-4}$ within 100k steps and apply 10k warm-up steps.

\subsection{SID}
\label{appendix_sid}

\paragraph{Dataset and Evaluation Metrics}
We use the official split of the VoxCeleb1 dataset \cite{nagrani2017voxceleb} for the SID task, where the test set contains 8,251 utterances from these 1,251 celebrities.
The capability of identifying speakers is assessed by classifying an utterance into the ground-truth category. Specifically, the whole utterance is taken as an input to the model to determine the speaker identity.

\paragraph{Fine-Tuning Details}
We use the cross-entropy loss and fine-tune all models on 8 V100 GPUs by the Adam optimizer with a batch size of 64 segments per GPU and the inputs of 3 seconds. The learning rate is set based on one cycle of a triangular cyclical schedule between $10^{-8}$ and $5\times10^{-4}$ in 60k steps. We initialize the weights of the text embeddings layer because there are no overlapping text tokens between the vocabularies during the pre-training and the SID fine-tuning.

\section{Results for 960 Hours Set of LibriSpeech}
\label{sec:appendix_960h}

We also fine-tune the model on the 960 hours set of LibriSpeech, as reported in Table \ref{exp_asr960}.
Experiments show that the proposed SpeechT5 model achieves significant improvement even without LM fusion, and it performs comparable or even better than wav2vec 2.0 with LM fusion.

\section{Results of the SpeechT5 without $\mathcal{L}_{mlm}^s$ on the TTS task}
\label{sec:appendix_tts_wo_hubert}

\begin{table}[!h]
\begin{center}
\small
\begin{tabular}{@{\extracolsep{5pt}}l|ccc@{}}
    \toprule
    % Model & WER & Naturalness \\
    Model & Naturalness \\
    % \midrule
    % Ground Truth & 4.1\% & - \\
    % HiFI-GAN & 5.1\% & - \\
    \midrule
    % SpeechT5 & 4.4\% & 2.79 \\
    % \hspace{2ex}w/o $\mathcal{D}^\text{S}$ & 3.8\% & 2.90 \\
    % \hspace{2ex}w/o $\mathcal{D}^\text{T}$ & 3.9\% & 2.77 \\
    % \hspace{2ex}w/o Joint PT & 4.2\% & 2.71 \\
    % \hspace{2ex}w/o $\mathcal{L}_{mlm}^{s}$ & 4.1\% & 2.91 \\
    SpeechT5 & 2.79 \\
    % \hspace{2ex}w/o $\mathcal{D}^\text{S}$ & 2.90 \\
    % \hspace{2ex}w/o $\mathcal{D}^\text{T}$ & 2.77 \\
    % \hspace{2ex}w/o Joint PT & 2.71 \\
    \hspace{2ex}w/o $\mathcal{L}_{mlm}^{s}$ & 2.91 \\
    \bottomrule
\end{tabular}
\end{center}
\caption{\label{tab:tts_as} Comparisons between SpeechT5 and its variant without using $\mathcal{L}_{mlm}^s$.}
\end{table}

We use the automatic evaluation tool NISQA-TTS to verify the performance of TTS results here, because it is convenient and cheap compared with MOS and CMOS, which need to be evaluated by humans.
As shown in Table \ref{tab:tts_as},
% the SpeechT5 trained by only text dataset achieves the optimal results in WER, and 
the variant of SpeechT5 trained without the loss $\mathcal{L}_{mlm}^{s}$ achieves an improvement in terms of naturalness when compared with the SpeechT5. 
% Since the WER of five variants are close to the groud truth speech, we consider the Naturalness as the main metric. 
% The variants without the speech dataset or the loss $\mathcal{L}_{mlm}^{s}$ are superior to the other variants, 
It suggests that the pre-training without 
% the speech dataset or 
the speech-specific loss brings a significant gain. Thus, we select the SpeechT5 without the loss $\mathcal{L}_{mlm}^{s}$ for MOS and CMOS evaluations.

\end{document}